\documentclass[prx,reprint,a4paper,superscriptaddress,citeautoscript,floatfix,svgnames]{revtex4-2}

\usepackage{amsmath,amsfonts,amssymb}
\usepackage{newtxtext,newtxmath}
\usepackage{mathrsfs}
\usepackage[pdftex]{graphicx}
\usepackage{microtype}
\usepackage{bm}\let\vec\bm

\usepackage[svgnames]{xcolor}
\usepackage[
	colorlinks=True,linkcolor=DarkRed,citecolor=ForestGreen,urlcolor=MediumBlue,
	pdfstartview=FitH,bookmarks=False,pdfpagemode=UseNone
]{hyperref}

\begin{document}
\draft


\title{Non-equilibrium Superconductivity and Quasiparticle Dynamics 
studied by Photo Induced Activation of Mm-Wave Absorption (PIAMA).}

\author{B. J. Feenstra$^{a,}$\cite{address}, J. Sch\"{u}tzmann$^a$, and D. van der Marel$^a$,
R. P\'{e}rez Pinaya$^b$, and M. Decroux$^b$}
\address{$^a$Materials Science Centre, Rijksuniversiteit Groningen,
Nijenborgh 4, 9747 AG Groningen, The Netherlands\\
$^b$D\'epartement de Physique de la Mati\`ere Condens\'ee, Universit\'e de 
Gen\`eve, 24 quai Ernest-Ansermet, CH-1211 Gen\`eve 4, Switzerland}

\date{May 8, 2018}
\begin{abstract}
We present a study of non-equilibrium superconductivity in
DyBa$_2$Cu$_3$O$_{7-\delta}$ using photo 
induced activation of mm-wave absorption (PIAMA). We 
monitor the time evolution of the thin film transmissivity 
at 5 cm$^{-1}$ subject to pulsed infrared radiation. 
In addition to a positive bolometric signal we observe a 
second, faster, decay with a sign opposite to the bolometric
signal for $T>40$ K. We attribute this to the unusual
properties of quasi-particles residing near the nodes of an
unconventional superconductor, resulting in a strong 
enhancement of the recombination time.
\end{abstract}

\maketitle



The occurrence of zero's in the superconducting gap 
for certain values of the momentum $\hbar\vec{k}$ at the 
Fermi surface of high T$_c$ superconductors has a number of
intriguing consequences for the dynamical behavior and life-time of 
the quasi-particles at low temperatures, which has only recently begun to
attract the attention of researchers in the field. Due to the presence of
these zero's (or nodes) the reduction in the superfluid fraction ($\rho _s$)
\cite{volovik,parks} and specific heat\cite{moler} is proportional to 
$H^{1/2}$, where $H$ is the magnetic field. 
Also, a strong reduction of the
quasi-particle scattering rate ($1/\tau $) below T$_c$ \cite{nuss,bonn,ong}
provides evidence that the dominant scattering mechanism has an electronic
signature.

In this Letter we present a study of the quasi-particle dynamics using Photo
Induced Activation of Mm-wave Absorption (PIAMA). In this pump/probe
experiment we use a free electron laser\cite{Felix} (FELIX) which is
continuously tunable from 100 to 2000 cm$^{-1}$ as a pump to create a
temporary excited state of a superconductor. 
FELIX produces macro-pulses with a stepwise off-on-off 
intensity profile ('on' for 3$<$t$<$7$\mu$s in Fig.\ 1), consisting of 5000
micropulses (1-5 ps). 
The step-response of the complex dielectric constant is monitored at 
5 cm$^{-1}$ using the combination of a Backward Wave Oscillator
(BWO) and a fast waveguide diode detector as a probe to measure the
transmission through a superconducting film as a function of time. The
mm-wave detector-circuit was selected as a compromise between sensitivity
and speed of detection, resulting in an overall time resolution of 1 $\mu$s.
This choice of experimental parameters is optimal for the detection of small
changes induced in the dielectric function by the infrared (IR) pulse at, as
we will see, the scale of the lifetime of nonequilibrium superconductivity
in high T$_c$'s.

We used films of DyBa$_2$Cu$_3$O$_{7-\delta }$ which were prepared by RF
sputtering on LaAlO$_3$ substrates. The film thickness was 20 nm and 
T$_c$ was 88 K. Optimal surface quality was obtained by using Dy instead
of Y. This substitution does not affect the superconducting properties.
A detailed description of the preparation, characterization and the mm-wave
dielectric properties of these films has been given elsewhere
\cite{perez,bjf1}.

\begin{figure}[htb]
\includegraphics[width=\columnwidth]{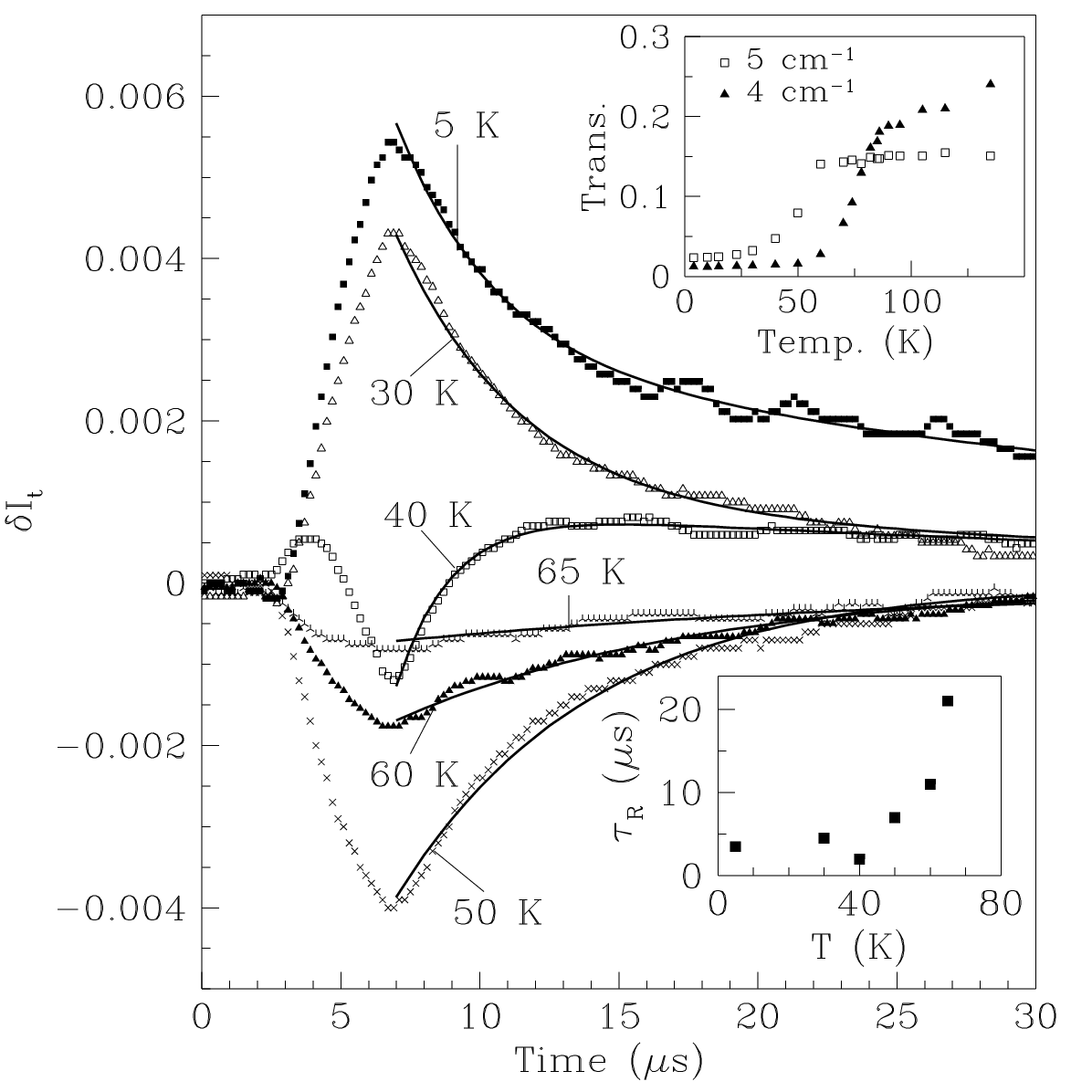}
\caption{{\protect\small Change in transmission, $\delta I_t$ for DyBa$_2$Cu$%
_3$O$_{7-\delta }$\hspace{0.1mm} on LaAlO$_3$, shown for several
temperatures. 
The FIR-pulse enhances transmission at low temperatures, while
it reduces it at temperatures higher than 40 K. The exponential fits are
shown as the solid lines. Inset, upper right corner: temperature dependence
of the unperturbed transmission at 4 and 5 cm$^{-1}$. Inset, lower right
corner: Temperature dependence of the faster relaxation time, $\tau _R$.}}
\label{tempdep}
\end{figure}

The LaAlO$_3$ substrate supporting the film is plan-parallel, with a
thickness $D=0.054\,{\rm cm}$ and a refractive index $n=4.70$. At $k/2\pi
\approx 5\,\mbox{cm}^{-1}$, which is our probe frequency, the dielectric
function of the film $|\epsilon |$ ranges from $10^4$ to $10^6$ depending on
temperature, while $(kd)^{-2}\approx 3\cdot 10^8$. Hence the films are
optically thin and $(kd)^{-2}\gg |\epsilon |\gg 1$. In that limit the
Fresnel expression for transmission through the film/substrate system is 
 \begin{equation}
 I_t=\left| 
   \left( 2-ikd\epsilon \right) \frac{\cos {\psi }}2
   -\left( i+in^2+kd\epsilon \right) \frac{\sin {\psi }}{2n}
  \right| ^{-2}
 \label{twotran}
 \end{equation}
where $\psi =nkD$. The effect of increasing the temperature is to transfer
spectral weight from the condensate to the quasi-particles, while at the
same time reducing the quasi-particle lifetime. The net result is, that both 
$|\epsilon ^{\prime }|$ and $\epsilon ^{\prime \prime }$ are reduced as the
temperature increases, and the thin film transmission increases. In the
inset of Fig.\ \ref{tempdep} the mm-wave transmission through a DyBa$_2$Cu$_3
$O$_{7-\delta }$~film of 20 nm thickness on a LaAlO$_3$ substrate is
displayed for $k/2\pi =5\,\mbox{cm}^{-1}$ and for $k/2\pi =4\,\mbox{cm}^{-1}$%
. The former corresponds to a larger sensitivity to the quasi-particles
(represented by $\epsilon ^{\prime \prime }$) as compared to the latter
frequency. A detailed analysis has been given elsewhere\cite{bjf1}. Most
significant for the identification of a possible bolometric response is the 
{\em monotonic} temperature dependence of the transmission over the entire
temperature interval.

\begin{figure}[htb]
\includegraphics[width=\columnwidth]{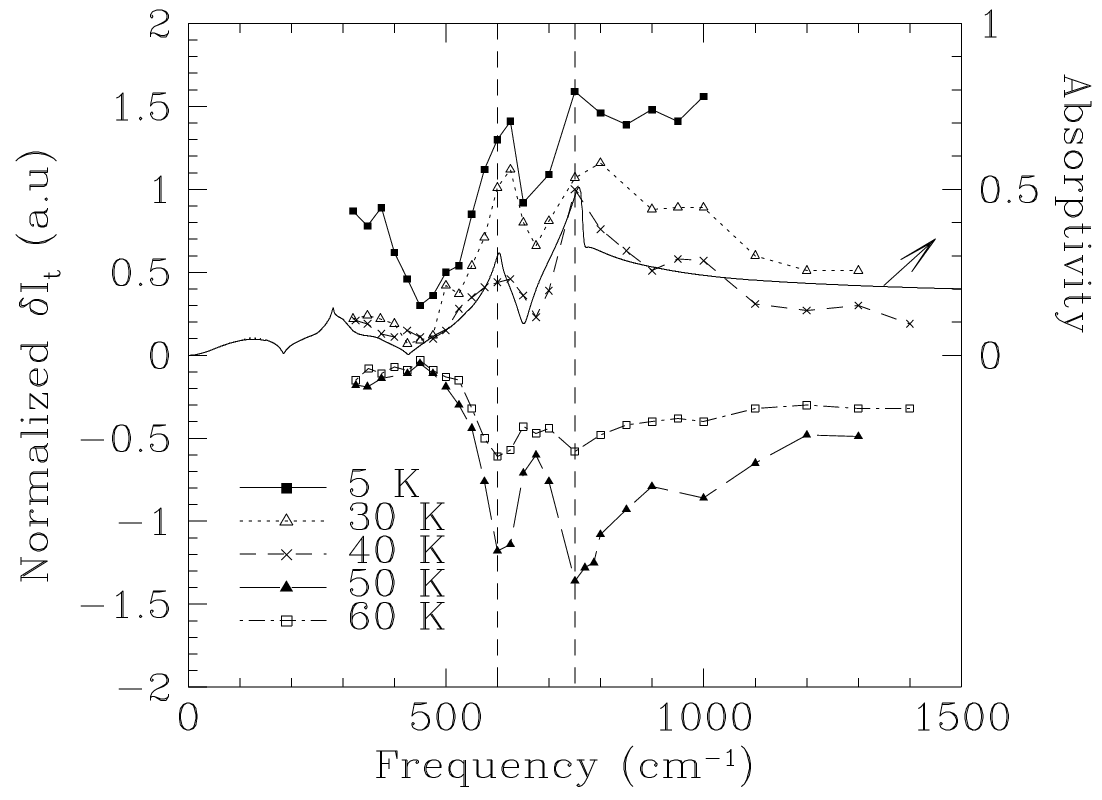}
\caption{{\protect\small Normalized $\delta I_t$ as a function of frequency,
shown for temperatures ranging from 5 to 60 K. Also shown is the
absorptivity within the film (solid line).}}
\label{pitpeak}
\end{figure}

In Fig.\ \ref{tempdep} the photo induced change in transmission ($\delta I_t$%
) of the same film is shown between 5 and 65 K. The probe frequency is $5\,%
\mbox{cm}^{-1}$. The pump frequency is $k/2\pi =800\,\mbox{cm}^{-1}$,
with a power of $\simeq$10 mJ/pulse. Here and in Fig.\ \ref
{pitpeak} the curves have been calibrated against variations in the incident
power of FELIX. We see that for temperatures lower than 40 K, the IR-pulse
enhances the transmissivity of the thin film. However, around 40 K the
situation changes and the transmission after the IR-pulse is reduced
instead. $\delta I_t$ is smaller at higher temperatures and becomes
undetectable above 75 K. The ordinary {\em monotonic} behavior seen in the
temperature dependence of the unperturbed mm-wave transmission indicates
that a simple heating of the sample can not account for the fact that $%
\delta I_t<0$ above 40 K. The fits shown in Fig.\ \ref{tempdep} correspond
to a linear combination of a slow ($\tau _B$) and a fast ($\tau _R$) decay: $%
\delta I_t=i_Be^{-t/\tau _B}+i_Re^{-t/\tau _R}$. The slow component $i_B$
has a weak time dependence ($\tau _B\gg 45\mu s$) on the interval displayed
in Fig.\ 1 and is reduced from $i_R/3$ at $5K$ to zero for $T>40K$. The
prefactor of the fast component ($4\mu s<\tau _R<25\mu s$) changes sign at
40 K. 
The changes in transmission as a function of pump frequency show a rather
non-monotonic behavior and have been summarized in Fig.\ \ref{pitpeak} for
several temperatures ranging from 5 to 60 K. Plotted are the maxima (minima)
of the positive (negative) peak intensities obtained after calibrating
against the changes in incident power. For comparison we display in the same
figure the absorption coefficient in the superconducting film $A_f=1-R_f-T_f$
of the IR-light, where $R_f$ is the reflectivity of the substrate-supported
film, and $T_f$ is the transmission through the film into the substrate. We
calculated $A_f$ without adjustable parameters from the experimentally
determined $a$- and $c$-axis dielectric function of YBaCuO \cite{marel,bauer}
and LaAlO$_3$ \cite{laalo3} using the Fresnel equations for light of mixed
polarization incident at an angle of 45$^{\circ }$ on a 20 nm thick, c-axis
oriented YBaCuO film on a LaAlO$_3$ substrate, identical to the experimental
situation. Optical absorption in the substrate occurs at 185, 427 and 651 cm$%
^{-1}$. For the film $A_f$ has minima at these frequencies and maxima at
290, 600 and 760 cm$^{-1}$, which is due to resonant reflection at the
substrate/film interface for frequencies matched to the longitudinal phonons
of the substrate.
The main conclusion from Fig.\ 2 is, that $\delta I_t$ tracks the laser power
deposited in the film, not in the substrate. This demonstrates that PIAMA
probes changes in the physical state of the superconductor, while secondary
effects due to substrate heating can be excluded.

The amplitude of the slow component, $\tau_B$, in Fig.\ \ref{tempdep}
corresponds to an increase in temperature of $13$ and $0.2$ K for the $5$ K
and $40$ K curves respectively. A crude estimate of the increase in
temperature based on the input laser-power and the specific heat of the
film/substrate system gives $\Delta T = 9$ and $0.2$ K respectively. We
therefore attribute the slow response to bolometric heating of the
film/substrate system. At higher temperatures the specific heat is too
large, and $\Delta T$ is insignificant. We observed a similar bolometric
response for MgO supported NbN thin superconducting films, in which case the
faster decay was absent within the limitations of the time resolution of our
detector. A more extensive discussion of this work is presented elsewhere%
\cite{bjf2}.

Let us now consider the faster decay, $\tau _R$. For sufficiently low
frequencies ($\omega \tau \ll 1$) the inductive response is proportional to
the condensate amplitude ($\epsilon ^{\prime }\propto \rho _s$), and will be
reduced during and following the IR-pulse, so that $\delta \rho _s<0$. Here
we are interested in the behavior of the quasi-particle response, which is
represented by the finite value of $\epsilon ^{\prime \prime }$. The latter
is proportional to the density of quasi-particles and their lifetime ($%
\epsilon ^{\prime \prime }\propto \rho _{qp}\tau $). Due to transfer of
spectral weight from the condensate to the quasi-particle peak we expect
that $\delta \rho _{qp}>0$ in the non-equilibrium state following the
IR-pulse. With PIAMA we attempt to probe the time-evolution of variations in
the volume density of quasi-particles ($\delta \rho _{qp}$). The highest
sensitivity to the latter variations relative to those of the condensate is
obtained for $\cos {\psi }=0$, which is also the experimental situation in
Figs.\ \ref{tempdep} and \ref{pitpeak}. In that case the transmission
coefficient varies as:
 \begin{equation}
  \frac{2n^2}{I_t^2}\delta I_t=-[kd\epsilon']^2\frac{\delta \rho _s}{\rho _s}
  -kd\epsilon''[1+n^2+kd\epsilon''] \frac{\delta \rho _{qp}\tau}{\rho _{qp}\tau}
 \label{deltran}
 \end{equation}
Immediately following the laser excitation the excess quasi-particles have
an enhanced non-equilibrium temperature $T^{*}$. During a short
time-interval of order 1 ns the quasi-particles thermalize to the
equilibrium temperature. {\em If} the quasi-particle recombination time ($%
\tau _R$) is long compared to the thermalization time, this leads to a
transient state with {\em cold} excess quasi-particles, where $\rho _{qp}$
is larger than the equilibrium value, but where $\tau $ is at its
equilibrium value, resulting in a $\delta I_t$ which is negative. Note that
this situation is different from ordinary heating of the sample, where the
quasi-particle peak also broadens ($\delta \tau <0$). The net-result in the
latter case is a reduction of $\epsilon ^{\prime \prime }$. The possibility
of cold excess quasi-particles was the subject of extensive investigations
in conventional superconductors\cite{Testardi,Rothwarf,Owen}.

The coefficients in Eq.\ (\ref{deltran}) are such\cite{bjf2}, that below 40K 
$\delta I_t$ is dominated by $\delta\rho_s$ (causing $\delta I_t>0$),
whereas above 40 K $\delta\rho_{qp}$ dominates (resulting in $\delta I_t<0$%
). Interestingly 40 K presents a bordercase, where $\delta I_t $ switches
sign from positive to negative when the pump-intensity exceeds a threshold
value. This observation is consistent with a series of experiments as a
function of incident laser power, pulse duration and pump frequency\cite
{bjf2}.

A lifetime of several $\mu$s for the nonequilibrium state is surprisingly
long compared to typical values reported for conventional superconductors,
ranging from 1 ns to 1 $\mu$s. Before discussing those considerations which
are specific to the high T$_c$ superconductors we recall that the most
important processes responsible for inelastic scattering are
electron-electron interactions and inelastic scattering by spin- and charge
fluctuations and phonons. At this point we want to stress that a net
decrease of the {\em number} of quasi-particles only results from phonon
assisted quasi-particle-pair recombination, {\em i.e.\ }events where {\em %
i.e.\ }two quasi-particles with momentum $\hbar k$ and $\hbar k^{\prime}$
and energy $E_k$ and $E_{k^{\prime}}$ are converted to a Cooper-pair and a
phonon with momentum $\hbar(k+k^{\prime})$ and energy $\Omega_{k+k^{\prime}}$%
. A similar conversion of quasi-particle pairs into electronic collective
modes may exist. However, as such modes can not escape into the substrate,
they will be converted back and forth into quasi-particles without reducing
the lifetime of the excited electron plasma.

In the cuprates several factors conspire to suppress the phonon-assisted
quasi-particle recombination processes. The vertex for phonon-assisted
quasi-particle recombination is the bare electron-phonon coupling constant ($%
g$) multiplied by the coherence factor $M_{kk^{\prime}}=u_{k^{%
\prime}}v_k+v_{k^{\prime}}u_k$. In isotropic $s$-wave superconductors the
lowest energy-levels accessible to a quasi-particle have an energy $\Delta$,
so that the quasi-particles are equally distributed along the Fermi-surface.
In a $d$-wave superconductor one expects quite different behavior: While
cooling down to an energy of order $k_BT$, the excess quasi-particles relax
toward the nodes. After this relaxation is completed, recombination
processes will only generate phonons with an energy of order $k_B T$ and
momentum of order $\hbar q_{ph}\approx k_BT/v_s$, where $v_s$ is the
sound-velocity. As $T \ll k_F v_s$, it follows that $q_{ph} \ll k_F$.
Hence most recombination processes will involve two quasi-particles in
nodes at opposite sides of the Fermi-surface. The coherence factor $%
M_{kk^{\prime}}$ is proportional to $\Delta_k/k_BT$, which becomes zero at
the nodes. Hence our first observation is that the quasi-particles relax to
those regions where the gap has its zero's, thus separating them from the
region in $k$-space with the largest pairing amplitude. This in turn leads
to a strong suppression of quasi-particle recombination processes. This
first observation is robust, and applies to general k-values of the nodes,
but also if zero's of the gap occur in coordinate space, {\em e.g.\ }in the
chain-bands, at defects, vortices {\em etc.}.

\begin{figure}[htb]
\includegraphics[width=\columnwidth]{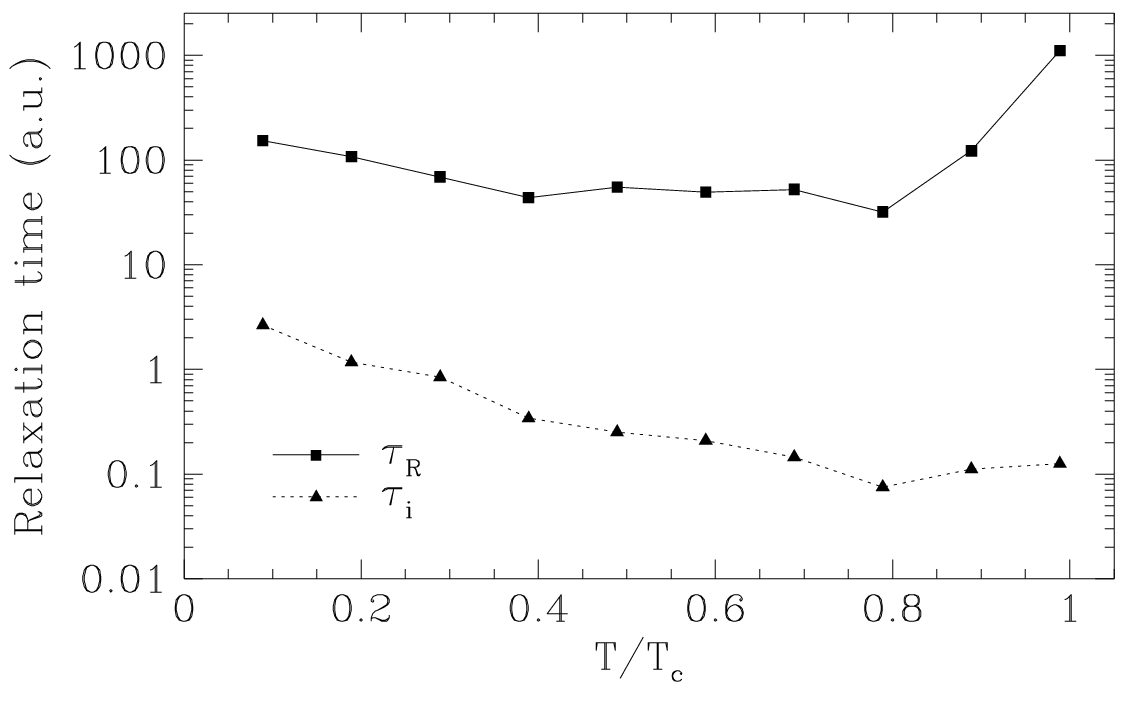}
\caption{{\protect\small Calculated temperature dependence of the
quasiparticle-phonon scattering time ($\tau _{in}$, triangles) and the
quasiparticle recombination time ($\tau _R$, squares).}}
\label{tmconst}
\end{figure}

For an isotropic $s$-wave superconductor energy conservation requires that
the recombination process is fully suppressed if $2\Delta$ exceeds the Debye
frequency. For a $d$-wave superconductor the situation is perhaps even more
intriguing. In thermal equilibrium the quasi-particles are concentrated near
the nodes. Near the nodes the two-dimensional energy-momentum relation 
has the functional form 
$E_k^2= (\partial_k\Delta)^2 k_t^2 + \hbar^2v_F^2 k_l^2$ 
where $k_t$ and $k_l$ are the momenta parallel
and perpendicular to the Fermi-surface measured relative to the node, and $%
\partial_k\Delta$ is the transverse momentum derivative of the
superconducting gap at the node. If $\partial_k\Delta$ (70 meV$\AA$ at 4 K%
\cite{schabel}) exceeds the sound-velocity ($v_s=$ 30 meV$\AA$ \cite
{reichardt}), the constraints on momentum and energy conservation can not be
satisfied, leading to a suppression of this process. Hence our second
observation is, that the phonon assisted quasi-particle recombination is
suppressed due to kinematical constraints when a large gap opens. An analogy
exists to the A-phase in superfluid $^3He$, where the relaxation rate of
quasi-particles near the nodes has an algebraic ($T^4$) temperature
dependence due to the reduction of the available phase space in the
superfluid phase\cite{wolfle}.
Together these arguments imply, that there is a strong suppression of
quasi-particle phonon scattering near the nodes, in particular of
quasi-particle recombination processes. 
We used Fermi's Golden Rule\cite{chang}
\begin{eqnarray}
\sum_k\frac{f_k}{\tau _R} &=&\sum_{k,k^{\prime }}g^2|M_{kk^{\prime }}|^2%
\mbox{Im}\frac{f_kf_{k^{\prime }}(1+n_q)\delta _{k^{\prime }}^{q-k}}{%
E_k+E_{k^{\prime }}-\Omega _q-i0^{+}}  \nonumber \\
\sum_k\frac{f_k}{\tau _i} &=&\sum_{k,k^{\prime }}g^2|L_{kk^{\prime
}}|^2\delta _{k^{\prime }}^{q+k}  \nonumber \\
&&\mbox{Im}\frac{f_k(1-f_{k^{\prime }})n_q+f_{k^{\prime }}(1-f_k)(1+n_q)}{%
E_{k^{\prime }}-E_k-\Omega _q-i0^{+}}  \label{tau}
\end{eqnarray}
to compute the thermal averages of the recombination lifetime $\tau _R$ and
the inelastic scattering time $\tau _i$ numerically. Here $f_k$ and $n_q$
are the Fermi-Dirac and Bose-Einstein distribution functions of
quasi-particles and phonons respectively, and $L_{kk^{\prime }}=u_{k^{\prime
}}v_k-v_{k^{\prime }}u_k$. We adopted a $d_{x^2-y^2}$ order parameter with $%
\Delta _{max}=25\,{\rm meV}$. The resulting temperature dependence of $\tau
_R$ and $\tau _i$ is displayed in Fig.\ \ref{tmconst}.
Most importantly we notice that $\tau_R\gg \tau_i$ at all temperatures.
The same calculation assuming an isotropic $s$-wave gap confirms the earlier
result that $\tau_R$ and $\tau_i$ are equal for $T\rightarrow T_c$ in $s$%
-wave superconductors\cite{Kaplan}.
A small admixture of s-wave symmetry of the type '$d$+$s$' merely breaks the 4-fold 
rotation symmetry of the quasiparticle dispersion, 
without affecting the physical picture.
With an admixture of the type '$d$+i$s$' the energy of the quasi particles near
the nodes is increased to $E_k^2=(\partial_k\Delta)^2
k_t^2 + \hbar^2 v_F^2 k_l^2 + \Delta_s^2$,
causing a further suppression of the available phase space for recombination 
processes, while $|M_{kk'}|$ increases near the nodes. The net effect 
on $\tau_R$ depends on $\partial_k\Delta$, $v_F$ and $\Delta_s$.

Finally we discuss our observations in relation to time-scales obtained with
other experimental techniques. Using micro-wave experiments the 
scattering time is found to change from 100 fs at 90 K to less than 10 ps at
40 K. With pump/probe experiments using visible light a decay of 0.2 ps has
been observed, which was associated with the life-time of quasiparticles
near the Fermi energy, along with a second slow decay of at least 20 ns\cite
{Stevens}. A relaxation of the resistivity within a few ns\cite{bluzer} has
been attributed to non-equilibrium quasiparticle-generation by hot phonons.
Based on an analysis of the critical flux-flow velocity Doettinger {\em et
al.}\cite{Doettinger} determined an inelastic scattering time ranging from
10 ps at 80 K to 0.1 $\mu$s at 40 K. The timescale of several $\mu$s
reported in this Letter is much longer. We attribute this to the fact that
the quasi-particle recombination time is always longer than the inelastic
scattering time, which is the sum of all electron-electron and
electron-phonon scattering processes, as is demonstrated by the numerical
calculation of the two electron-phonon time constants $\tau_R$ and $\tau_i$
presented above.

In conclusion, we have observed a non-equilibrium state with a life-time of
several $\mu$s in DyBa$_2$Cu$_3$O$_{7-\delta}$\hspace{0.1mm} using photo
induced activation of mm-wave absorption (PIAMA). The non-equilibrium state
is clearly distinct from bolometric heating of the superconductor. The long
time-constant seems to reveal an unusually long quasi-particle recombination
time, which can be understood as a consequence of the highly peculiar nature
of quasi-particles near the nodes in these materials. Along with other
factors, such as the amplitude of the gap, the presence of nodes
distinguishes these materials from conventional superconductors.

{\em Acknowledgements} We gratefully acknowledge the assistance by
the FELIX staff, in particular A. F. G. van der Meer. Furthermore we thank
W. N. Hardy, D. I. Khomskii and O. Fisher for their stimulating comments
during the preparation of this manuscript and A. Wittlin for fruitfull
discussions at the initial stage of this project.


\begin{references}
\vspace{-1.5cm}
\bibitem[*]{address} Present address: Center of Superconductivity 
Research, Univ.\ of Maryland, College Park, MD 20742-4111
\bibitem{volovik}   G.E. Volovik, JETP Lett. {\bf 58}, 469 (1993).

\bibitem{parks}   B. Parks {\em et al.}, 
Phys. Rev. Lett. {\bf 74}, 3265 (1995).

\bibitem{moler}   K. A. Moler {\em et al.}, 
Phys. Rev. Lett. {\bf 73}, 2744 (1994).

\bibitem{nuss}   M. C. Nuss {\em et al.}, 
Phys. Rev. Lett. {\bf 66}, 3305 (1991).

\bibitem{bonn}   D. A. Bonn {\em et al.}, 
Phys. Rev. Lett. {\bf 68}, 2390 (1992).

\bibitem{ong}   K. Krishana, J. M. Harris, and N. P. Ong, Phys. Rev. Lett. 
{\bf 75}, 3529(1995).

\bibitem{Felix}   G. M. Knippels {\em et al.}, 
Phys. Rev. Lett. {\bf 75}, 1755 (1995).

\bibitem{perez}   
Roberto P\'{e}rez {\em et al.},
IEEE Trans. Appl. Superconductivity {\bf 7}, in press (1997).

\bibitem{bjf1}   B. J. Feenstra {\em et al.}, 
Physica C {\bf 278}, 213 (1997).

\bibitem{marel}   D. van der Marel {\em et al.}, 
Phys. Rev. B {\bf 43}, 8606 (1991).

\bibitem{bauer}  R. Gajic {\em et al.}, 
J. Phys. Condens. Matter {\bf 4}, 1643 (1992). 

\bibitem{laalo3}   Z. M. Zhang {\em et al.}, 
J. Opt. Soc. Am. B {\bf 11}, 2252 (1994).

\bibitem{bjf2}   B. J. Feenstra, Ph D thesis, Univ.\ of Groningen (1997).

\bibitem{Testardi}   L. R. Testardi, Phys. Rev. B {\bf 4}, 2189 (1971).

\bibitem{Rothwarf}   A. Rothwarf, G. A. Sai-Halasz and D. N. Langenberg,
Phys. Rev. Lett. {\bf 33}, 212 (1974).

\bibitem{Owen}   C. S. Owen and D. J. Scalapino, Phys. Rev. Lett. {\bf 28},
1559 (1972).

\bibitem{schabel}   Matthias C. Schabel {\em et al.}, Phys. Rev. B {\bf 55},
2796 (1997).

\bibitem{reichardt}   W. Reichardt {\em et al.}, Physica C {\bf 162-164},
464 (1989).

\bibitem{wolfle}   D. Vollhardt and P. W\"{o}lfle, {\em The Superfluid
Phases of Helium 3}, Tayler \& Francis, London, New York, Philadelphia
(1990).

\bibitem{chang}   J.-J. Chang, in {\em Nonequilibrium Superconductivity,
Phonons and Kapitza boundaries}, Kenneth E. Gray ed. (Plenum Press, New
York, 1981).

\bibitem{Kaplan}   S. B. Kaplan {\em et al.}, 
Phys. Rev. B {\bf 15}, 3567 (1977).

\bibitem{Stevens}   C. J. Stevens {\em et al.}, 
Phys. Rev. Lett. {\bf 78}, 2212 (1997).

\bibitem{bluzer}   N. Bluzer, Phys. Rev. B, {\bf 44}, 10222 (1991).

\bibitem{Doettinger}   S. G. Doettinger {\em et al.}, 
Phys. Rev. Lett. {\bf 73}, 1691 (1994).
\end{references}
\end{document}